\documentclass[twocolumn,superscriptaddress,showpacs,prd,aps,amsmath,amssymb,nofootinbib]{revtex4-1}
\usepackage{amsmath, amsfonts, hyperref}
\usepackage{graphicx}
\newcommand{\beq}{\begin{equation}}
\newcommand{\eeq}{\end{equation}}
\newcommand{\beqn}{\begin{eqnarray}}
\newcommand{\eeqn}{\end{eqnarray}}

\newcommand{\apjl}{Astrophys.\ J.\ Lett.}
\newcommand{\mnras}{Mon.\ Not.\ R.\ Astro.\ Soc.}

\begin{document}
\title{
Star Clusters, Self-Interacting Dark Matter Halos and
Black Hole Cusps: \\ The Fluid Conduction Model and its Extension to 
General Relativity
}
\date{\today}
\author{Stuart~L.~Shapiro}
\altaffiliation{Also Department of Astronomy and NCSA, University of
  Illinois at Urbana-Champaign, Urbana, Illinois 61801, USA}
\affiliation{Department of Physics, University of Illinois at
  Urbana-Champaign, Urbana, Illinois 61801, USA}

\begin{abstract}
We adopt the fluid conduction approximation to study the
evolution of spherical star clusters and self-interacting dark matter (SIDM)
halos. We also explore the formation and dynamical impact of 
density cusps that arise in both systems due to the presence of a
massive, central black hole. The large N-body, self-gravitating
systems we treat are ``weakly-collisional": the mean free time 
between star or SIDM particle collisions is much longer
than their characteristic crossing (dynamical) time scale, but shorter than 
the system lifetime. The fluid conduction model reliably tracks the
``gravothermal catastrophe" in star clusters and SIDM halos without black holes.
For a star cluster with a massive, central black hole, this approximation 
reproduces the familiar Bahcall-Wolf 
quasistatic density cusp for the stars bound to the black hole
and shows how the cusp halts the ``gravothermal catastrophe" and causes the 
cluster to re-expand. 
An SIDM halo with an initial black hole central density spike that matches 
onto to an exterior NFW profile relaxes to a core-halo 
structure with a central density cusp determined by the
velocity dependence of the SIDM interaction cross section.
The success and relative simplicity of the fluid conduction approach in
evolving such ``weakly-collisional", quasiequilibrium Newtonian systems
motivates its extension to relativistic systems. We present a
general relativistic extension here.

\end{abstract}
\maketitle

\section{Introduction}
\label{sec:intro}

The fluid conduction approximation has been adopted successfully
to study the dynamical evolution of a spherical star cluster, 
(see, e.g.,~\cite{LynE80,Spi87, BetS84,BetI85,Goo87,HegR89,BalSI02,ShaP14})
as well as a self-interacting dark matter (SIDM) halo 
(see, e.g. ~\cite{BalSI02,AhnS05,ShaP14}).
In this approach the ensemble of 
gravitating particles is modeled by a ``weakly-collisional" 
fluid in quasistatic, virial equilibrium.
The local temperature is identified with the square of the
velocity dispersion and thermal heat conduction is employed to
reflect the manner in which orbital motion and scattering combine
to transfer energy in the system. The basis of the heat conduction equations
are moments of the Boltzmann equation, substituting a simple model for
the collision terms.

This gravothermal fluid formalism was originally introduced for the study
of globular star clusters, where it has proven to be very 
useful in understanding the {\it secular} evolution of these 
systems on relaxation timescales. The agreement between this fluid 
approach with more detailed (e.g, Fokker-Planck) treatments 
comes about despite the fact that star clusters are only 
weakly-collisional and have long collision mean free paths greatly 
exceeding the size of the cluster, where thermalization is 
achieved by the cumulative effect of repeated, distant, 
small-angle gravitational (Coulomb) encounters. The Fokker-Planck
equation treats the phase space distribution function $f$, whose evolution 
is driven by diffusion coefficients involving integrals of $f$ over
the entire system. By contrast, the fluid conduction equations evolve
locally defined quantities (the density and velocity dispersion at a given
spatial coordinate), although this approach incorporates a 
relaxation timescale and an effective mean free path in the heat conductivity 
that are based on global considerations and 
collision integrals over the entire system.  
In fact, the fluid conduction description 
may be even better suited to SIDM halos, for which the dominant
thermalizing particle interactions in some 
models may be close-encounter, 
large-angle (hard-sphere) scatterings. It is reassuring, nevertheless, 
that, even in the case of weakly-collisional systems such
as star clusters, the fluid conduction prescription does reproduce 
many of the results found in more fundamental
analyses of the (weakly) collisional Boltzmann equation, with collisions 
treated via more precise, but computationally more demanding, 
Monte Carlo approaches or direct Fokker-Planck integrations 
(see reviews in, e.g., ~\cite{Spi87,LigS78,Sha85,ElsHI87,BinT87}, 
and a recent summary of methods in ~\cite{Vas15}, and references therein).
All of these approaches can be extended to treat anisotropic
and multicomponent systems.

An isolated, self-gravitating N-body system 
in virial equilibrium will relax via gravitational encounters
(scattering) to a state consisting of an extended halo 
surrounding a nearly homogeneous, isothermal central core. As time advances,
the core transfers mass and
energy through the flow of particles and heat to the extended halo.
The thermal evolution timescale of the dense core is much shorter than that
of the extended halo, which essentially serves as a quasistatic heat sink. 
As the core evolves it shrinks in size and mass, while its density 
and temperature grow. Increase of central temperature induces 
further heat transfer to the halo, leading to a {\it secular}
instability on a thermal (collisional relaxation)
timescale. The secular contraction of the core towards infinite density and
temperature but zero mass is known as the 
``gravothermal catastrophe''(see, e.g.,~\cite{Spi87}). 
The late-time, homologous nature of this secular 
instability is well-described by the fluid conduction model, as first shown by 
Lynden-Bell and Eggleton~\cite{LynE80}. They solved the equations by 
separation of variables, looking for 
a self-similar solution applicable at late times.

A critical departure from the secular contraction scenario occurs when
a {\it dynamical} instability sets in, which can occur when
the particle velocities in the core, or, equivalently, when the central 
potential, become relativistic. 
As originally conjectured by Zel'dovich and Podurets~\cite{ZelP66} and
explicitly demonstrated by Shapiro and 
Teukolsky~\cite{ShaT85a,ShaT85b,ShaT86,ShaT92}, 
collisionless systems in virial equilibrium typically experience a radial
instability to collapse on dynamical timescales when their cores become
sufficiently relativistic. This dynamically instability 
terminates the epoch of secular gravothermal contraction in clusters 
and leads to the catastrophic collapse of a core of {\it finite} mass 
to a black hole. The general relativistic 
simulations of the catastrophic collapse of relativistic
clusters, which are essentially collisionless on dynamical timescales, 
by Shapiro and Teukolsky were performed in part to explore the
possible origin of the supermassive black holes (SMBHs) 
that exist at the centers of most galaxies and quasars.
Such a SMBH formation scenario might occur in relativistic 
clusters of compact stars following the gravothermal 
catastrophe~{\cite{Ree84,ShaT85c,QuiS87,QuiS89}. 
A similar SMBH formation scenario may also occur in SIDM halos, as
originally proposed by Balberg and Shapiro~\cite{BalS02}. 

The existence of dense clusters of stellar-mass black holes and/or other
compact objects in the cores of galaxies has been given a boost by the
recent discovery of a swarm of black holes, inferred to be $2 \times 10^4$ 
in number, within one parsec of the supermassive black hole 
Sagittarius A* at the center of the Galaxy~\cite{HaiMBBHH18}.
Concentrations ranging from several thousands to tens of thousands 
of stellar-mass black holes in this region have long been predicted by
numerous investigators (see,e.g.,~\cite{Mor93,MirG00,FreAK06,GenSMO18}).
Such systems in the nuclei of other galaxies 
have been suggested (see, e.g.,~\cite{ElbBK18,AntR16} and references 
therein) as the likely sites for the formation of the
binary black holes whose mergers have been observed by 
Advanced LIGO/VIRGO (e.g., ~\cite{Abb16,Abb17}).

Here we briefly review the fluid conduction model and solve it numerically
to evolve spherical, isotropic, single component 
star clusters and SIDM halos. To calibrate our code, we first integrate
the full system of equations, starting from a Plummer model, to 
track the full evolution and development of the gravothermal catastrophe 
in star clusters. 
By contrast, the original treatment using this approach
for isolated star clusters (see, e.g., ~\cite{LynE80} and the summary
in ~\cite{Spi87}) only considered the late-time, self-similar behavior, 
after the gravothermal catastrophe was well underway.
We then apply the model to probe the effect of a massive, central black hole
on the cluster density and velocity profiles, 
and its impact on the secular evolution of the system. 
While these features have been studied previously, they have
not been analyzed by solving the fluid conduction equations.
We recover the familiar Bahcall-Wolf (~\cite{BahW76}, hereafter BW) 
power-law profiles 
that dominate the central cusp embedded in a {\it static}, nearly homogeneous, 
isothermal core of equal-mass stars: the cusp density varies with radius as
$\rho \propto r^{-7/4}$ and the velocity varies as $v \propto r^{-1/2}$. 
We then show how the presence of the cusp, which drives heat into core, 
eventually halts the gravothermal catastrophe, causing it to reverse its
contraction and the cluster to re-expand. We predicted this
behavior using a simple homologous cluster 
model~\cite{Sha77} and it was corroborated subsequently by 
our Monte Carlo simulations of the two-dimensional Fokker-Planck equation for 
the stellar phase space distribution function $f(E,J;t)$ describing
a spherical cluster containing a central black hole~\cite{MarS80,DunS82}
(see also~\cite{Vas17}).

In this paper we next apply the fluid conduction model to
isolated SIDM halos, following up on our original treatment~{\cite{BalSI02} 
of these systems. We previously explored the gravothermal 
catastrophe in such systems, probing both the late self-similar 
evolution of a typical system in which the mean free path 
between collisions $\lambda$ is initially 
longer than the scale height $H$ everywhere (which is always true in 
a star cluster), and then
tracking the  general time-dependent evolution of such systems. 
We found that $\lambda$ can eventually become smaller than $H$ in the 
innermost core, at which point that region behaves like 
a conventional fluid. At 
late times the core becomes relativistic and likely unstable
to dynamical collapse to a black hole, as discussed above. 
We then determined the 
steady-state cusp that forms around a massive central black
in an ambient, {\it static} SIDM core, solving the steady-state 
fluid conduction equations both in 
Newtonian gravity and general relativity~\cite{ShaP14}. 
We showed that the density in the cusp scales with radius
as $r^{-\beta}$ for an interaction cross section that varies 
with velocity as 
$\sigma \sim v^{-a}$, where $\beta=(a+3)/4$.

By contrast, here we allow an SIDM halo to 
evolve in response to the central black hole. 
Specifically, we consider an SIDM halo born with a 
Navarro-Frenk-White (NFW)~\cite{NavFW97} density profile by the usual 
collisionless, cosmological halo formation mechanism. We 
assume that soon thereafter a central density spike forms 
in the halo in response to the adiabatic growth of a massive, 
central black hole. 
We then show, by solving the fluid conduction equations,
that the spike evolves into a BW-like cusp which drives heat into the
ambient halo, ultimately causing the core to expand, as in the case of a
star cluster.

Given the utility of the fluid conduction model as demonstrated anew by the
above applications, we present for the first 
time the full set of fluid conduction 
equations for following the secular evolution of
a weakly-collisional system in general relativity. General relativistic
simulations have been performed for the dynamical evolution
of completely collisionless systems, as summarized above, as well as 
for relativistic fluid systems, such as stars. But as far as we are
aware, there have been no implementations of a scheme to track 
the secular evolution of relativistic systems that are ``weakly-collisional".  
Yet as discussed above, 
the gravothermal catastrophe in star clusters and SIDM halos can 
ultimately drive Newtonian systems to a ``weakly-collisional" relativistic 
state. The secular evolution of such a relativistic system 
immediately thereafter is governed neither by the 
collisionless Boltzmann (Vlasov) equation nor the ``strongly-collisional" 
equations of relativistic hydrodynamics.
It is thus necessary to provide a general relativistic formalism to bridge
the epochs from Newtonian secular core contraction to relativistic 
dynamical collapse, and we do so here.

We emphasize that by focusing on the fluid conduction model 
in this paper we in no way offer it as a substitute for
the more precise approaches mentioned above
that have been designed, at least in Newtonian theory, 
to track the detailed evolution of weakly-collisional, 
large $N$-body, self-gravitating systems.
Rather, our treatment here is presented to highlight the robustness and
versatility of a scheme that is capable of physically 
reliable, first approximations to solutions of a great 
many problems that can be obtained with a minimum of computational 
resources and time. All calculations reported in this paper 
were performed on a single laptop. 
In the case of relativistic systems, 
we provide an approach where no schemes have been presented previously. 

In Sec. II we present the Newtonian fluid conduction equations
for spherical, isotropic systems and cast them into two different 
forms, both of which are  useful  
numerically. In Sec. III we apply these equations to probe 
the secular evolution of several 
astrophysically realistic systems. These 
include a star cluster that begins as a Plummer
model and undergoes the gravothermal catastrophe, as well as a Plummer model
in which we suddenly insert a massive, central black hole and follow the
resulting dynamical behavior. We also 
treat the secular evolution of SIDM halos containing a 
massive, central black hole. 
In Sec IV we present the general relativistic fluid conduction equations for
spherical, isotropic systems. 

We adopt geometrized units and set $G=1=c$ throughout.

\section{Newtonian Treatment}
\label{sec:Newt}

The basic Newtonian fluid conduction equations 
are given by~\cite{LynE80, Spi87, BalSI02, ShaP14}
\begin{equation}
\label{eq:mass}
\frac{\partial M}{\partial r} = 4 \pi r^2 \rho
\end{equation}

\begin{equation}
\label{eq:hydroeq}
\frac{\partial (\rho v^2)}{\partial r}=-\rho~ \frac{M + M_h}{r^2} 
\end{equation}

\begin{eqnarray}
\label{eq:firstlaw}
\frac{\partial L}{\partial r} &=& - 4\pi r^2 \rho 
\left\{\frac{D}{Dt}\frac{3 v^2}{2}+
      P\frac{D}{Dt}\frac{1}{\rho}\right\}  \cr \cr \cr
&=& 
-4\pi r^2 \rho v^2 \frac{D}{Dt}
\ln\left(\frac{v^3}{\rho}\right)  
\end{eqnarray}

Equation~(\ref{eq:hydroeq}) 
is the equation of hydrostatic equilibrium,
where $\rho$ is the matter density, $v$ is the one-dimensional matter 
velocity dispersion, $M=M(r)$ is the mass of matter 
interior to radius $r$, 
$M_h$ is the central black hole mass, if present, and 
$P$ is the the kinetic matter pressure, which satisfies $P=\rho v^2$. 
Equation ~(\ref{eq:firstlaw}) is the the first law of
thermodynamics for the rate of change of $\ln s$, the specific 
entropy of the matter, where we define $s$ by 
\begin{equation}
\label{eq:entropy}
s=\left(\frac{v^3}{\rho}\right).
\end{equation}
The quantity $L$ is the luminosity due to heat conduction.
The time derivatives in Eq.~(\ref{eq:firstlaw}) are Lagrangian, 
and follow a given mass element. 

For all applications considered in this paper $L$ is a 
conductive heat flux evaluated in the {\it long} mean free path 
limit,
\begin{equation}
\label{eq:flux}
\frac{L}{4 \pi r^2}=-3 b \rho \frac{H^2}{t_r} 
          \frac{\partial v^2}{\partial r}.
\end{equation}
(But see~\cite{BalSI02}, Eq.~13 for the more general case).
In writing Eq.~(\ref{eq:flux}) we evaluated
the kinetic temperature of the particles according to 
$k_B T = mv^2$, where $k_B$ is Boltzmann's constant.
The parameter $b$ is constant of order unity 
and $H$ is the local particle
scale height. The quantity $t_r$ is the local relaxation timescale.
Its functional form depends on the matter interactions (Coulomb
scattering for stars, other possibilities for SIDM) and will be assigned
below for each application. 

In the absence of a massive, central black hole the scale height is 
taken to be the local Jeans length $H=r_J = (3v^2/12 \pi \rho)^{1/2}$ 
(see, e.g., \cite{Spi87}, Eq. 1-24) By contrast, in the presence of a 
black hole, the matter at $r$ that is bound to the black hole in the cusp and 
moves in a potential dominated by the hole has a scale height 
that is comparable to its characteristic orbital radius $r << r_J$.
For a system containing a black hole it proves sufficient then
to set $H = {\rm min}(r, r_J)$, which
accommodates the matter both inside and outside the cusp.

It is straightforward to generalize the set of equations to accommodate
multicomponent systems containing particles of different masses and/or 
species.
In such cases there will be separate hydrostatic equilibrium and entropy evolution 
equations for each component. In each entropy equation there will be, 
in addition to the self-interaction heat conduction term,  
pairwise thermal coupling terms to all the other components. 
These terms are each proportional to the difference in the 
local temperatures of the components 
and conduct heat from hotter to colder 
members (see, e.g., ~\cite{BetI85}). The effect of these coupling terms is to drive the system to 
equipartition, which in turn leads to mass segregation. 
In this paper, however, we shall focus on single component systems.

It is sometimes 
computationally useful to express the evolution equations using 
$M=M(r)$ as the independent Lagrangian variable in order to maintain
adequate coverage of the matter over the vast dynamical range of 
density and radius that accompanies the gravothermal instability or 
the formation of a cusp around a central black hole. Consequently we have
$r=r(t,M)$, $\rho = \rho(t,M)$, etc, and 
Eqs.~(\ref{eq:mass})-(\ref{eq:flux}) become
\begin{equation}
\label{eq:mass2}
\frac{\partial r}{\partial M} = \frac{1}{4 \pi r^2 \rho},
\end{equation}

\begin{equation}
\label{eq:hydroeq2}
\frac{\partial (\rho v^2)}{\partial M} = -\frac{M + M_h}{4 \pi r^4} \rho,
\end{equation}

\begin{equation}
\label{eq:flux2}
\frac{L}{4 \pi r^2}=-3 b \rho \frac{H^2}{t_r} 
          4 \pi \rho r^2 \frac{\partial v^2}{\partial M},
\end{equation}

and

\begin{equation}
\label{eq:firstlaw2}
\frac{D}{Dt} \ln \left(\frac{v^3}{\rho} \right) =
-\frac{1}{v^2} \frac{\partial L}{\partial M}.
\end{equation}

The above system of equations for a virialized cluster 
is quite similar in form to the
equations of stellar evolution, where one is also solving for the
secular evolution of a configuration in hydrostatic equilibrium.

\subsection{Star Clusters}

\subsubsection{Relaxation Timescale}

In a star cluster relaxation is driven by multiple,
small-angle gravitational (Coulomb) encounters. The local relaxation
time scale is given by (see e.g., \cite{LigS78,Spi87})

\begin{eqnarray}
\label{eq:tr_stars}
t_r({\rm stars}) &=&  
\frac{3^{3/2} v^3}{4 \pi \alpha m \rho \ln(0.4 N)},  \cr
&\simeq& 0.7 \times 10^{9}\mbox{yr} 
\left(\frac{v}{\mbox{km}~{\mbox sec}^{-1}}\right)^3 \cr
&\times& \left(\frac{M_{\odot}{\mbox{pc}^{-3}}}{\rho}\right)
\left(\frac{M_{\odot}}{m}\right)
\left( \frac{1}{\ln(0.4N)}\right),
\end{eqnarray}
where $\alpha = 1.22$, $m$ is the stellar mass and
$N$ is the total number of stars in the cluster. 

\subsubsection{Nondimensional Equations}

It is computationally convenient to cast the fluid conduction
Eqs.~(\ref{eq:mass2})-(\ref{eq:firstlaw2}) 
into nondimensional form. This is accomplished by introducing a fiducial
mass $M_0$ and radius $R_0$, in terms of which corresponding nondimensional 
parameters are denoted by a tilde according to
\begin{eqnarray}
\label{eq:scale}
r &=& R_0 {\tilde r}, \cr
M &=& M_0 {\tilde M}.
\end{eqnarray}
The parameters $M_0$ and $R_0$ then define a characteristic 
velocity, density, timescale, and entropy parameter,
\begin{eqnarray}
\label{eq:nondim}
v_0 &=& \left( \frac{M_0}{R_0} \right)^{1/2}, \;\;\; 4 \pi \rho_0 = 
\left( \frac{M_0}{R_0^3} \right), \cr
t_0 &=& t_{r0} \frac{1}{6b} ~, \qquad\qquad\, s_0 = \frac{v_0^3}{\rho_0} ~,
\end{eqnarray}
which yield corresponding nondimensional quantities,
\begin{eqnarray}
\label{eq:nondim2}
v &=& v_0 {\tilde v}, \quad \rho = \rho_0 {\tilde \rho_0}  
\quad t = t_0 {\tilde t} \quad s = s_0 {\tilde s}.
\end{eqnarray}
In Eq.~(\ref{eq:nondim}) $t_{r0}$ is the relaxation timescale in
Eq.~(\ref{eq:tr_stars}), evaluated for  $v=v_0$ and $\rho=\rho_0$. 
The parameter $b$ appearing in Eqs.~(\ref{eq:flux2}) and ~(\ref{eq:nondim})
is equal to $0.45$ for star clusters~\cite{Spi87}. 
Inserting Eq.~(\ref{eq:flux2}) into Eq.~(\ref{eq:firstlaw2}), 
writing Eqs.~(\ref{eq:mass2})-(\ref{eq:firstlaw2}) in terms of
nondimensional variables, and then dropping the tildes, yields
\begin{equation}
\label{eq:mass3}
\frac{\partial r}{\partial M} = \frac{1}{r^2 \rho}
\end{equation}

\begin{equation}
\label{eq:hydroeq3}
\frac{\partial (\rho v^2)}{\partial M} = -\frac{M + M_h}{r^4} \rho
\end{equation}

\begin{equation}
\label{eq:firstlaw3}
\frac{D}{Dt}{\ln}~s = \frac{1}{v^2} \frac{\partial}{\partial M}
\left[ r^4 \rho^2 \left( \frac{H^2}{r_J^2} \right) 
\frac{\partial v}{\partial M} \right].
\end{equation}

An alternative form for the entropy equation is
\begin{equation}
\label{eq:firstlaw4}
\frac{D s}{Dt} =
\frac{v}{\rho} \frac{\partial}{\partial M} \left[ \frac{r^4 \rho^3}{3 v^2} 
\left( \frac{H^2}{r_J^2} \right) \frac{\partial s}{\partial M} 
+ \frac{r^4 \rho v}{3} \left( \frac{H^2}{r_J^2} \right) 
\frac{\partial \rho}{\partial M} \right].
\end{equation}

The optimal way of numerically integrating the fluid conduction equations
presumably would be to implement the Henyey method, as in a typical,
battle-tested, stellar evolution code 
(see a description in, e.g., \cite{Cla68}, Sec 6-4).
In the interest of obtaining quick results with minimal 
code writing or adaptation, it has proven adequate to integrate
Eqs.~({\ref{eq:mass3})-~(\ref{eq:firstlaw3}) via a straightforward
{\it explicit}  forward-time, center-spaced finite-difference scheme. 
First, the diffusion-like (parabolic)  evolution
equation~(\ref{eq:firstlaw3}) is integrated
forward in time on a timestep $\Delta t$ restricted by the
(crudely estimated) Courant timestep: 
\begin{eqnarray}
\label{eq:Courant}
\Delta t &=& 0.5 ~{\rm min} \left[ 
\frac{(\Delta M)^2}{D} \right] \times C, \cr \cr
D &\approx& \frac{r^4 \rho^2}{3 v} \left( \frac{H^2}{r_J^2}\right),
      \ \ \ C \approx \mathcal{O}(1),  
\end{eqnarray}
where $\Delta M$ is the grid spacing, $D$ is an effective diffusion 
constant, $C$ is a constant Courant factor of order unity,
and the minimum is taken over all the grid points.

Next, using the value of $s(t,M)=v^3/\rho$ obtained at the new time, 
Eqs.~(\ref{eq:mass3}) and ~(\ref{eq:hydroeq3})
are iterated to obtain $r(t,M), \rho(t,M)$ and $v(t,M)$ on that time.
Solving Eq.~(\ref{eq:firstlaw3}) in a follow-up predictor-corrector step (or 
adopting a higher-order, time-centered, iterative scheme) is a refinement that 
was tested but proven unnecessary in practice for reliable results. 
The spatial differencing is second order in the Lagrangian variable $M$.

For clusters containing black holes, tracking the very late evolution
and re-expansion proves difficult with the above explicit scheme, as 
the Courant timestep plummets when the cusp develops and the central
(Lagrangian) grid spacing drops as the central density grows. Instead, we integrate 
Eq.~(\ref{eq:firstlaw4}) rather than Eq.~(\ref{eq:firstlaw3}), grouping the
terms linear in $s$ and evaluating $s$ {\it implicitly} in time. 
Solving the resulting linear (tridiagonal) finite-difference 
equations for $s$ is no longer governed by a Courant condition for 
stability, so longer timesteps tuned to
the evolution timescale, and not the decreasing conduction timescale across  
a central grid point, can be exploited (i.e. $C$ can be chosen much larger
than unity).
Once $s$ is determined on the new timestep, Eqs.~(\ref{eq:mass3}) 
and ~(\ref{eq:hydroeq3}) may be iterated as before.

\subsubsection{The Gravothermal Catastrophe}
\label{sec:gravo}

As our first application we track the secular evolution of a cluster
that begins as a Plummer model without a central black hole 
(i.e. $M_h = 0$).

\paragraph{Initial Data.}

A Plummer model is an equilibrium polytrope of index $n=5$ that has
a finite total mass $M_P$ and an infinite radius. We cut off
the cluster at a finite radius containing 99\% of the total mass. 
The Plummer density, velocity and mass profiles are given by 
(see, e.g., ~\cite{Spi87}, Eqs.~(1-17)-(1-19))
\begin{eqnarray}
\label{eq:Plum}
\rho(r)&=&\frac{3M_P}{4 \pi a^3} ~\frac{1}{\left(1+r^2/a^2\right)^{5/2}} \cr
v^2(r)&=&\frac{M_P}{a} ~\frac{1}{6 \left(1+r^2/a^2\right)^{1/2}} \cr
M(r)&=&M_P ~\frac{r^3/a^3}{\left(1+r^2/a^2\right)^{3/2}}
\end{eqnarray} 
We use the total mass $M_P$ and the scale factor $a$ 
to set the mass and radius scale introduced in Eq.~(\ref{eq:scale}):
$M_0 \equiv M_P$ and $R_0 \equiv a/2^{1/2}$.

\paragraph{Boundary Conditions}

We assume regularity at the cluster center, e.g.,
\begin{equation}
\label{eq:bcin}
\frac{\partial \rho}{\partial r} \to 0, \ \ \ 
\frac{\partial v}{\partial r} \to 0, \ \ \ \ 
M \to 0,
\end{equation}
and take the density and pressure to vanish at the
surface,
\begin{equation}
\label{eq:bcout}
\rho = 0, \ \ \ \rho v^2 = 0, \ \ \ M = 1.
\end{equation}
These conditions suffice to determine the system 
and are implemented in the finite difference
equations. For example, Eqs.~(\ref{eq:bcin}) and (\ref{eq:bcout}) 
are both used in finite differencing Eqs.~(\ref{eq:firstlaw3}) 
and ~(\ref{eq:firstlaw4}),
while Eq.~(\ref{eq:bcout}) is used in Eq.~(\ref{eq:hydroeq3}),
starting at the cluster surface and integrating inward.

\paragraph {Numerical Results}

The fluid conduction system of equations were finite-differenced
with 281 grid points in $M$, logarithmically spaced. We set $H=r_J$ everywhere. 
The evolution equation for $s$ was integrated in time both explicitly 
via Eq.~(\ref{eq:firstlaw3}) and implicitly via Eq.~(\ref{eq:firstlaw4}).
In both cases the timestep was set by Eq.~(\ref{eq:Courant}) 
with $C$ equal to 3, although considerably higher values of $C$
also proved satisfactory using the implicit version, as expected.
The two sets of integrations gave results that were very comparable; 
we will describe those obtained with the explicit implementation below.
Integrations with half and twice as many grid points showed convergence with
decreasing grid spacing.

The results of the numerical integration are summarized in 
Figs.~\ref{fig:dens_Plum}--~\ref{fig:gravo_Plum}. The asymptotic behavior
revealed in the plots at late times clearly exhibits
the familiar gravothermal instability in a star cluster. 
Fig.~\ref{fig:dens_Plum}
plots snapshots of the density profile at selected times and shows
that the nearly homogeneous core undergoes contraction on a secular
timescale, growing in central density while encompassing an ever 
decreasing fraction of the total mass. Once the contraction is well underway
($t \gg t_{rc}(0)$, where $t_{rc}(0)$ is the initial central relaxation
timescale) the density approaches the
self-similar solution of Lynden-Bell and Eggleton~\cite{LynE80} for 
gravothermal collapse.
In particular, the density profile in the envelope approaches
$\rho \propto r^{-(2+\beta)} \propto r^{-2.21}$, where 
$\beta = (1-\zeta)/(2-\zeta)$ and $\zeta = 0.737$~(cf.~\cite{Spi87}, 
Eqs. (3-33),(3-34)). Fig.~\ref{fig:vel_Plum} plots snapshots of the 
velocity dispersion 
profile at corresponding times, showing that the shrinking core is 
nearly isothermal, while the envelope dispersion 
scales as $v \sim (M(r)/r)^{1/2} \sim
(\rho r^2)^{1/2} \propto r^{-0.11}$. Fig.~\ref{fig:gravo_Plum} illustrates
good agreement with the asymptotic  
temporal relations that characterize the asymptotic self-similar
solution~(cf.~\cite{Spi87}, Eqs.~(3-6)--(3-8), (3-46) and (3-47)):
\begin{eqnarray}
\label{eq:collap}
\frac{\rho_c}{\rho_c(0)} &=& \left[1-\frac{t}{t_{\rm coll}}\right]^
    \frac{-2(5-3\zeta)}{(7-3\zeta)} = 
    \left[1-\frac{t}{t_{\rm coll}}\right]^{-1.165}, \cr \cr \cr 
\frac{v_c}{v_c(0)} &=& \left[ \frac{\rho_c}{\rho_c(0)}\right]^
    \frac{(1-\zeta)}{2(5-3\zeta)} =
    \left[1-\frac{t}{t_{\rm coll}}\right]^{-0.0550},
\end{eqnarray}
where $t_{\rm coll}$ is the core collapse time, at which the central
density $\rho_c$ blows up to infinity while the core mass shrinks to zero.
The velocity is thus seen to change much more
slowly than the density during the collapse.
We also recover asymptotically the self-similar solution 
result that the time remaining before
complete collapse is a constant multiple of the instantaneous central
relaxation time (cf. \cite{Spi87}, Eq. (3-47)),
\begin{equation}
\frac{t_{\rm coll}-t}{t_{rc}} = \frac{2(5-3\zeta)}{(7-3\zeta)}
      \frac{1}{\xi_c} \approx 320,
\end{equation} 
where $\xi_c \approx 3.6 \times 10^{-3}$.
\begin{figure}
\includegraphics[width=10cm]{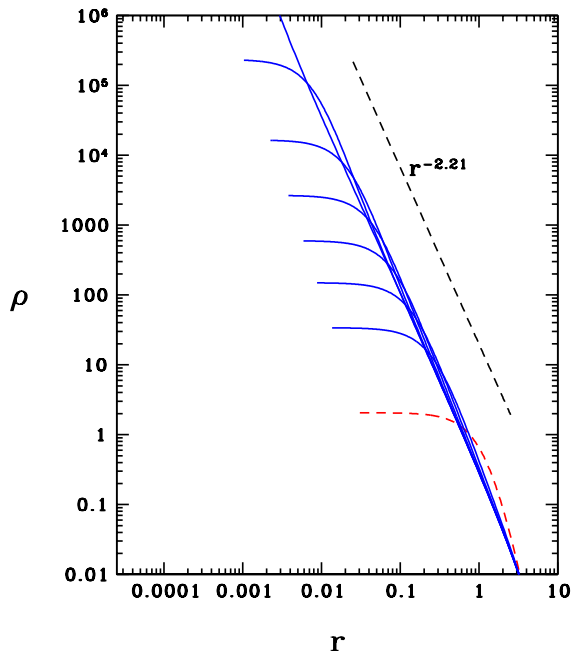}
\caption{The gravothermal castastrophe: 
snapshots of the density profile at selected times.
The curved {\it dashed} red line shows the density at time $t=0$.
Successively higher {\it solid} blue curves show the density at
$t =$ 2.107, 2.826, 3.159, 3.313, 3.378. 3.399, and 3.402. 
The straight {\it dashed} black 
line shows the slope for the self-similar solution, to which the
envelope asymptotes at late times.  All quantities are
in nondimensional units defined in Eqs.~(\ref{eq:scale}--\ref{eq:nondim2}),
for which $t_0 \approx 9.71 t_{rc}(0)$.
}
\label{fig:dens_Plum}
\end{figure}

\begin{figure}
\includegraphics[width=10cm]{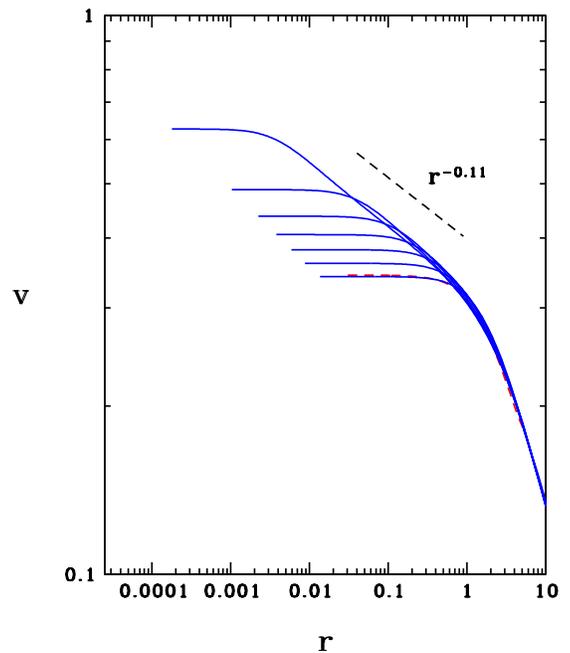}
\caption{The gravothermal catastrophe: 
snapshots of the velocity dispersion profile at the same selected times
depicted in Fig.~\ref{fig:dens_Plum}. The core velocity dispersion 
increases with time.
The straight {\it dashed} black 
curve shows the slope for the self-similar solution, to which the
envelope asymptotes at late times. All quantities are
in nondimensional units defined in Eqs.~(\ref{eq:scale}--\ref{eq:nondim2}).
}
\label{fig:vel_Plum} 
\end{figure}

\begin{figure}
\includegraphics[width=10cm]{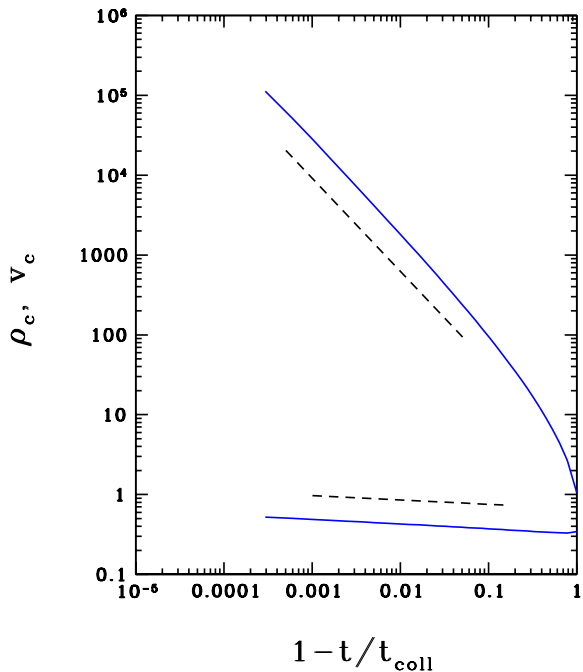}
\caption{The gravothermal catastrophe: 
the central density (upper two lines) and velocity dispersion 
(lower two lines) as functions of time. The {\it solid} blue lines 
plot the results of integrating the full fluid conduction equations, while the
{\it dashed} black lines show the late-time 
self-similar slopes, Eq.~\ref{eq:collap}. The quantity $t_{\rm coll}$ is
the core collapse time. All quantities are
in nondimensional units defined in Eqs.~(\ref{eq:scale}--\ref{eq:nondim2})
}
\label{fig:gravo_Plum}
\end{figure}

\subsubsection{Black Hole in a Static Ambient Cluster}
\label{subsub:static}

Here we probe the formation of the cusp around a massive
black hole $M_h << M_P$ inserted at the center of the same 
Plummer star cluster
described in Eq.~(\ref{eq:Plum}).
We {\it fix} for all $t \geq 0$ the cluster profile outside the 
inner core but allow the region near and within the black hole's
zone of influence at $r \leq r_h$ to relax in the presence of the hole. 
Here $r_h=M_h/v^2(0)$, where $v(0)$ is
the central velocity dispersion in the initial cluster. 
The velocity dispersion is nearly constant in the core and remains 
unperturbed well outside $r_h$.  We expect the cluster to
evolve to the BW profile in the cusp and relax to a steady-state.
Once again we take $M_0 \equiv M_P$ and $R_0 \equiv a/2^{1/2}$

\paragraph{Initial Data}

We take $r_h/r_{\rm core} = 10^{-3}$, 
where the core radius $r_{\rm core}$ is defined to be the radius at
which the density $\rho$ falls to one-half its central value:
$r_{core} = a (2^{2/5}-1)^{1/2}$ or $r_{core} = 0.799$ in our nondimensional
units. This choice of $r_h$ gives 
the black hole a mass $M_h/M_P = 0.942 \times 10^{-4}$,
By construction, $M_h$ is much less than the total mass $M_P$ of the stars in 
the cluster, but much greater than the mass within the cusp ($r < r_h$),  
both initially and after steady-state is reached
($M(r_h)/M_h = 4.60 \times 10^{-6}$ at late times). 
Accordingly, the gravitational potential of the black hole dominates 
that of the stars in the cusp. This is the regime modeled by BW.
We take the same Plummer density profile given in Eq.~(\ref{eq:Plum}) 
but we solve Eq.~(\ref{eq:hydroeq2}) for the initial velocity 
dispersion to ensure that the cluster with the central black hole is 
in virial equilibrium at the start of its secular evolution in
the core. We neglect any initial contribution within $r_h$ from 
stars unbound to the black hole. They will generate a weak
$r^{-1/2}$ cusp ~\cite{ShaT83} that will be swamped by the cusp that
forms from the bound stars as they begin to relax.

\paragraph{Boundary Conditions}

An ordinary star of radius $R$ and mass $m$ 
is tidally disrupted by the black hole 
whenever it passes within a radius $r_D$, where
\begin{equation}
r_D \sim R(M_h/m)^{1/3}
\end{equation}
However, sufficiently compact stars, such as neutron stars or
stellar-mass black holes, may avoid tidal disruption before reaching 
the the marginally bound radius $r_{mb}$, where
\begin{equation}
r_{mb} = 4 M_h
\end{equation}
in Schwarzschild coordinates. Even a main sequence star like the sun
would escape disruption if the black hole exceeds $\sim 10^8 M_{\odot}$. 
Any star that penetrates within
$r_{mb}$ must plunge directly into the black hole (see, e.g, the discussion 
in \cite{ShaP14} and references therein).
To mimic either scenario we fix a small inner radius $r_{in}$ within 
which the interior stellar mass is set to a vanishingly small value.
\begin{equation}
\label{eq:bcinfixed}
r = r_{in}, \ \ \  M \to 0, 
\end{equation}
which implies $\rho = 0$ for $r < r_{in}$.
We put $r_{in}/r_h = 3.81 \times 10^{-2}$ to illustrate the effect.

The outer boundary $r_{out}$ is taken well outside the black hole 
radius of influence but 
well inside the core radius:  
$r_{out} = 11.1 r_h = 1.11 \times  10^{-2} r_{core}$.
At $r_{out}$  we match all quantities to the Plummer model
parameters, which are held fixed during the evolution:
\begin{equation}
r=r_{out}, \ \ \, \rho=\rho_P, \ \ \, \rho v^2 = \rho_P v_{P}^2, 
    \ \ \ M = M(r_{out}).
\end{equation}
With these assignments 
$M(r_{out})/M_P=2.51 \times 10^{-7}$
and $M(r_{core})/M_P=0.119$. 

\begin{figure}
\includegraphics[width=10cm]{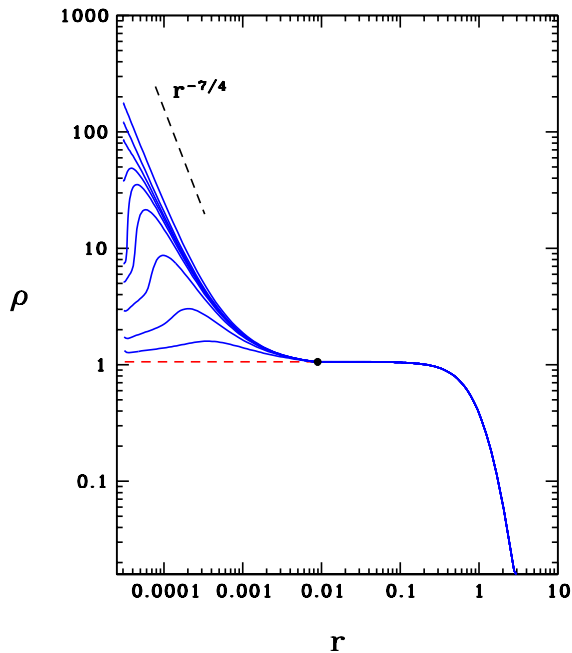}
\caption{Black hole cusp in a static core:
snapshots of the density profile at selected times.
The lower {\it dashed} red line shows the density at time $t=0$.
Successively higher {\it solid} blue curves show the density at
$t =$ 0.0764, 0.1366, 0.1862, 0.2121, 0.2241, 0.2316, 0.2395, 0.2537,
and 0.3330.  The solid {\it dot} indicates
the matching radius $r_{out}$, outside of which the profile is held fixed.
The upper {\it dashed} black
line shows the slope for the steady-state BW solution, to which the
cusp relaxes.  All quantities are
in nondimensional units defined in Eqs.~(\ref{eq:scale}--\ref{eq:nondim2}),
for which $t_0 \approx 9.71 t_{rc}(0)$.
}
\label{fig:dens_BHfixedbc}
\end{figure}

\begin{figure}
\includegraphics[width=10cm]{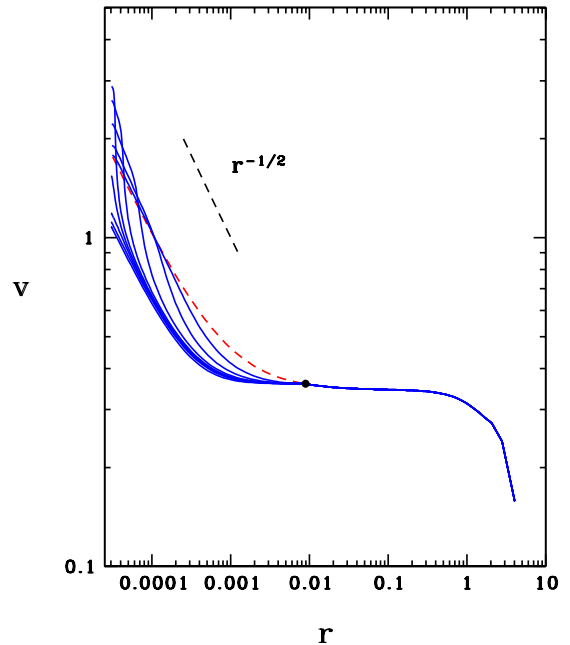}
\caption{Black hole cusp in a static core: 
snapshots of the velocity dispersion profile at the same selected times
depicted in Fig.~\ref{fig:dens_BHfixedbc}. Successively lower 
{\it solid} curves are at increasing time.
The solid {\it dot} indicates
the matching radius $r_{out}$, outside of which the profile is held fixed.
The straight {\it dashed} black
curve shows the slope for the steady-state BW solution, to which the
cusp relaxes.  All quantities are
in nondimensional units defined in Eqs.~(\ref{eq:scale}--\ref{eq:nondim2}).
}
\label{fig:vel_BHfixedbc}
\end{figure}

\paragraph{Numerical Results}

The system of equations was integrated with 141 grid points covering 
the cluster, but with only 75 points inside $r_{out}$. The explicit
form of the entropy evolution equation, Eq.~(\ref{eq:firstlaw3}),
with Courant factor $C=1$ proved adequate. The evolution of the 
density and velocity profiles
are shown in Figs.~\ref{fig:dens_BHfixedbc} and ~\ref{fig:vel_BHfixedbc}.

Relaxation drives the cusp to the familiar steady-state, power-law BW profile, 
as anticipated.  Removing the constraint that the interior match to a
fixed cluster core at $r_{out}$ will allow the cluster to evolve, as we
will see in the next section. The BW solution for the cusp is readily seen as 
a consequence of the fluid conduction equations 
in steady state, in which case $L(r)=constant$ independent of $r$, 
according to Eq.~(\ref{eq:firstlaw}). We used this result
previously~\cite{ShaL76} to derive the BW density profile from simple scaling.
Now, by setting $\rho \propto r^{-\beta}$ and $M_h \gg M$
in Eq.~(\ref{eq:hydroeq}) we obtain $v^2 \approx [1/(\beta +1)]M_h/r$ 
inside the cusp.  Requiring steady-state in Eq.~(\ref{eq:firstlaw}
gives $L=constant$, which when  
inserted into Eq.~(\ref{eq:flux}) with $H \sim r$ 
yields $\beta = 7/4$, as found by BW. The numerical integrations
are in good agreement with these steady-state profiles.

\subsubsection{Black Hole in an Evolving Cluster}
\label{subsub:BHevol}

Here we begin with the same cluster and central black hole
as in Section~\ref{subsub:static} above, but now we remove all
constraints and allow the cluster to evolve. 
We are interested in observing the competition between
those encounters that lead to the
the gravothermal catastrophe and drive secular core collapse
versus those arising from heating by the black hole cusp and drive 
core expansion. 

\paragraph{Initial Data} We adopt the same initial data as in the previous
section.

\paragraph{Boundary Conditions} We adopt the same black hole-induced 
inner boundary condition as in the previous section, 
Eq.~(\ref{eq:bcinfixed}). The outer boundary for such an isolated,
freely-evolving cluster is set at the cluster surface 
via Eq.~(\ref{eq:bcout}).  
\begin{figure}
\includegraphics[width=10cm]{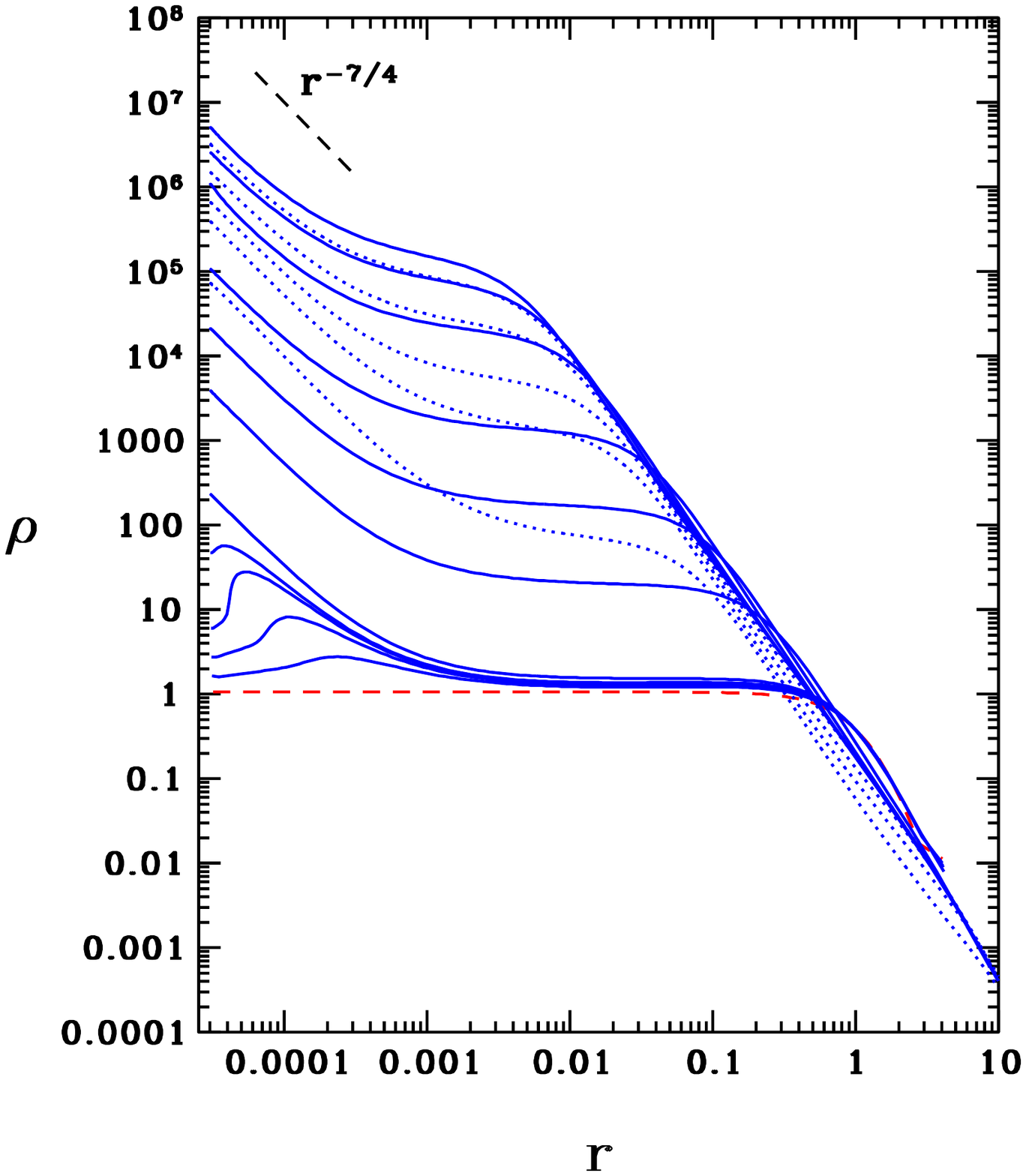}
\caption{Black hole influence on cluster evolution:
snapshots of the density profile at selected times.
The lower {\it dashed} red line shows the density at time $t=0$.
Successively higher {\it solid} blue curves show the density at
$t = 0.1286, 0.1844, 0.2163, 0.2306, 0.3447, 5.954,
9.611, 10.98, 11.44, \newline 
11.48$ and $11.51$ (gravothermal collapse). Successively lower
{\it dotted} blue curves then show the density at $t = 11.53, 11.81,
14.36, 23.11$ and $62.50$ (re-expansion).  The upper {\it dashed} black
line shows the slope for the steady-state BW cusp solution.
All quantities are in nondimensional units defined in
Eqs.~(\ref{eq:scale}--\ref{eq:nondim2}),
for which $t_0 \approx 9.71 t_{rc}(0)$.
}
\label{fig:dens_BHplum_nospike}
\end{figure}

\begin{figure}
\includegraphics[width=10cm]{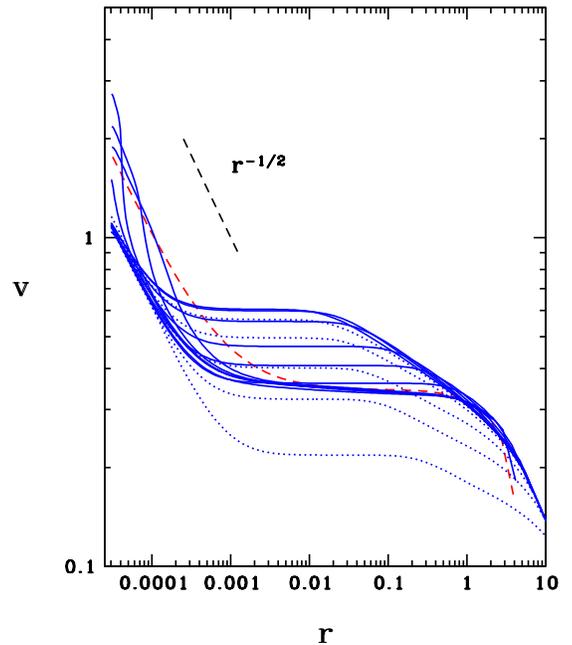}
\caption{Black hole influence on cluster evolution:
snapshots of the velocity dispersion profile at the same selected times
depicted in Fig.~\ref{fig:dens_BHplum_nospike}.
The straight {\it dashed} black
curve shows the slope for the steady-state BW cusp solution, 
to which the cusp quickly relaxes. Once the dispersion velocity 
adjusts to the density cusp, the {\it solid}
curves show successively higher core dispersions 
with increasing time (gravothermal collapse). 
The {\it dotted} curves then follow, showing successively lower
core dispersions with increasing time (re-expansion). All 
quantities are
in nondimensional units defined in Eqs.~(\ref{eq:scale}--\ref{eq:nondim2}).
}
\label{fig:vel_BHplum_nospike}
\end{figure}

\paragraph{Numerical Results} We employ a grid of 141 points to integrate
the system of equations, using the entropy evolution in the form given by
Eq.~(\ref{eq:firstlaw4}) and solving it implicitly. A variable Courant 
constant $C$ was chosen for the timestep set by Eq.~(\ref{eq:Courant}),
increasing from $C=5$ at early times to $C=2 \times 10^{12}$ at late times.  
The key reasons for the increase in $C$ are the huge growth in $\rho$ with time at
the inner boundary of the cusp and the fact that $\Delta t$ as given by
Eq.~(\ref{eq:Courant}) plummets like $\rho^{-3}$ in this region.

The evolution of the cluster is summarized in 
Figs.~\ref{fig:dens_BHplum_nospike}
and ~\ref{fig:vel_BHplum_nospike}. The early evolution in the cusp 
for $t \lesssim 0.23$ proceeds much as did in the previous application, 
where the ambient cluster was held fixed beyond the outer core.
During this epoch the cusp, where the relaxation timescale is shortest, 
evolves in response to the presence of the black
hole and again approaches a BW profile. But during an intermediate
evolutionary phase when $0.23 \lesssim t \lesssim 11.5$ the cluster interactions
trigger incipient gravothermal core collapse. During this epoch
the cusp maintains a BW profile with a density that smoothly matches onto the
ever-increasing core density just outside $r_h$. The late evolution
when $t \gtrsim 11.5$ is characterized by secular core re-expansion.
Heating from the cusp drives the expansion, causing the core density and 
velocity disperion to fall and the outer mass shells to increase in radius.
We predicted such expansion from a simple homologous cluster model in
~\cite{Sha77} and probed its detailed nature by solving the
Fokker-Planck equation by Monte Carlo simulations in ~\cite{MarS80,DunS82}
(see also~\cite{Vas17}). 
It is reassuring to see that the fluid conduction approach
recovers the same qualitative behavior when a massive black hole resides
at the center of a cluster.

\subsection{Self-Interacting Dark Matter}

We have applied the fluid conduction approximation to track the
secular evolution of isolated SIDM halos in previous studies.
Our initial application~\cite{BalSI02} treated the secular gravothermal
catastrophe in Newtonian halos subject to elastic,
velocity-independent interactions. There we showed that 
in typical halos $\lambda$, the mean free path for scattering, is much larger 
than the scale height $H$ initially, but once the contracting core evolves to 
sufficiently high density, the inequality is reversed 
in the innermost regions. This central region then behaves like a hydrodynamic 
fluid core surrounded by a weakly-collisional halo. We 
suggested~\cite{BalS02} that black hole formation is an inevitable
consequence of the gravothermal catastrophe in SIDM halos 
once the core becomes sufficiently relativistic, as it becomes
radially unstable and undergoes dynamical collapse. This scenario
may produce the massive seed black holes that later merge 
and accrete gas to become the supermassive black holes 
observed in most galaxies and quasars.

We returned to the subject recently when we applied
the fluid conduction approximation to model the steady-state
distribution of matter around a massive black hole at the center 
of a weakly-collisional SIDM halo~\cite{ShaP14}. There we
allowed the interactions to be governed by a velocity-dependent
cross section $\sigma \sim v^{-a}$, solved the steady-state equations
both in Newtonian theory and general relativity and
showed that the SIDM density in the cusp scales as $ \rho \sim r^{-\beta}$ away
from the cusp boundaries, where
$\beta = (a+3)/4$, while its velocity dispersion satisfies 
$v^2 \approx [1/(\beta+1)]M_h/r$ or 
$v \sim r^{-1/2}$. For $a=4$ the interaction cross section 
has the same velocity dependence as Coulomb scattering and 
we recover the BW profile. In this case the solution 
we found applies to stars in a star cluster as well as SIDM. These
steady-state calculations assumed that the ambient halo outside the
cusp remained static, as we did in Section~\ref{subsub:static} above 
for star clusters.

Missing from the above steady-state analysis is a time-dependent
calculation that shows how the SIDM density and velocity profiles 
secularly evolve away from their initial configurations. Those
initial configurations likely
include a central density spike that arises 
early on, following the appearance and adiabatic growth of a 
central supermassive black hole on timescales shorter than
the dark matter self-interaction relaxation time, $t_r(SIDM)$. 
The spike then evolves on the timescale $t_r(SIDM)$ into a 
weakly-collisional cusp and the entire halo then expands 
in response to the heat driven into the halo by the cusp. 
We shall perform a simulation that illustrates this behavior below.

\subsubsection{Relaxation Timescale}

In a SIDM halo relaxation is driven by elastic interactions 
between particles.  The relaxation time scale
is the mean time between single collisions and is given by
\begin{eqnarray}
\label{eq:tr_sidm}
t_r({\rm SIDM}) &=& \frac{1}{\eta \rho v \sigma} \cr
&\simeq& 0.8\times 10^{9}\mbox{yr}
             \left[\left(\frac{\eta}{2.26}\right)
             \left(\frac{\rho}{10^{-24}\mbox{g cm}^{-3}}\right)\right. \cr
 &\times& \left.\left(\frac{v_*}{10^7\mbox{cm sec}^{-1}}\right)
\left(\frac{v}{v_*}\right)^{1-a}
       \left(\frac{\sigma_0}{1~\mbox{cm}^2\;\mbox{g}^{-1}}\right)\right]^{-1}\,
\end{eqnarray}
where $\sigma = \sigma_0 (v/v_*)^{-a}$ is the cross section per unit mass and
the constant $\eta$ is of order unity. For example,
$\eta = \sqrt{16/\pi}\approx 2.26$ for particles
interacting elastically like billiard balls (hard spheres) with a
Maxwell-Boltzmann velocity distribution
[see \cite{Rei65}, Eqs.~(7.10.3), (12.2.9) and (12.2.13)]. 
\footnote{For a brief discussion of some cosmologically and physically
viable choices  for $\sigma_0$ and $a$ see \cite{ShaP14} and 
references therein.} We note again that
for a Coulomb-like cross section, where $a=4$,  $t_r({\rm SIDM})$ scales
the same way with $v$ and $\rho$ as
$t_r({\rm stars})$:
$t_r \propto v^3/\rho$. 

\subsubsection{Nondimensional Equations}

We modify the scalings for $\rho_0$ and $t_0$ defined 
in Eqs.~(\ref{eq:nondim}) and (\ref{eq:nondim2}) by
choosing instead
\begin{equation}
\label{eq:nondim3}
\rho_0 = \left( \frac{M_0}{R_0^3} \right), \;\;\;  
t_0 = t_{r0} \frac{1}{6b}\frac{1}{4 \pi},
\end{equation}
while keeping the other scalings the same. 
Here $t_{r0}$ is given by Eq.~(\ref{eq:tr_sidm}), evaluated
for $v=v_0$ and $\rho=\rho_0$. 
For a gas of hard spheres with a Maxwell-Boltzmann distribution
the coefficient $b$ in Eq.~(\ref{eq:flux}) 
can be calculated to good precision from transport theory,
and has the value of $b\approx (25/64)\sqrt{2\pi/3} \approx 0.565$
[cf. \cite{LifP81}, Sec. 10, Eq.~(7.6) and Problem 3, and \cite{Spi87}, Eq.~(3-35)]. For a gas
obeying a Coulomb scattering cross section 
$b \approx 0.45$~\cite{Spi87}.
The resulting nondimensional 
equilibrium Eqs.~(\ref{eq:mass3})
and ~(\ref{eq:hydroeq3}) are unchanged
but the entropy evolution Eq.~(\ref{eq:firstlaw3}) now becomes
\begin{equation}
\label{eq:firstlaw5}
\frac{D}{Dt}\ln s = \frac{1}{v^2} \frac{\partial}{\partial M}
\left[ r^4 \rho^2 v^{4-a} \left( \frac{H^2}{r_J^2} \right) 
\frac{\partial v}{\partial M} \right],
\end{equation}
while Eq.~(\ref{eq:firstlaw4}) becomes
\begin{equation}
\label{eq:firstlaw6}
\frac{D s}{Dt} =
\frac{v}{\rho} \frac{\partial}{\partial M} \left[ \frac{r^4 \rho^3}{3 v^{a-2}} 
\left( \frac{H^2}{r_J^2} \right) \frac{\partial s}{\partial M} 
+ \frac{r^4 \rho v^{5-a}}{3} \left( \frac{H^2}{r_J^2} \right) 
\frac{\partial \rho}{\partial M} \right].
\end{equation}

The above equations apply to the long mean-free path (LMFP) limit that 
characterizes the early and longest secular evolution phase of an
SIDM halo and the phase we wish to probe here. 
For the more general equations that handle the transition
from the early LMFP phase to the late, short mean free path (SMFP) phase, when
such a transition occurs, see ~\cite{BalSI02}. 

\subsubsection{The Gravothermal Catastrophe}

We previously treated in Ref~\cite{BalSI02}
the evolution of a SIDM halo in the absence of a black hole 
and with a velocity-independent ($a=0$) interaction cross-section
using the fluid conduction equations, and we will not repeat the 
analysis here. There we showed how a halo can evolve from the
(self-similar) LMFP regime to the SMFP regime in the inner core of the
halo and discussed how the catastrophic collapse of the core can
naturally provide the seed for a supermassive black hole at the halo
center. We discussed this SIDM-black hole formation scenario in greater detail
in Ref~\cite{BalS02}. 

\subsubsection{Black Hole in a Static Ambient Cluster}
\label{subsub:static2}

As mentioned in Sec.~\ref{sec:intro}, this scenario was treated in Ref~\cite{ShaP14},
both in Newtonian and general relativistic gravitation.
We took $a=4$ for the velocity dependence in the 
SIDM interaction cross section in the example we worked out.
We noted that any depletion in the DM density deep in the spike due to
DM annihilation~\cite{GonS99,Vas07,ShaS16} would be washed out by 
self-interactions.
We refer the reader to that paper for further details.

\subsubsection{Black Hole in an Evolving Halo}
\label{subsub:BHevol2}

Here we consider the full evolution of a SIDM 
halo, formed in the early Universe with an 
NFW profile, that houses 
a massive seed black hole at its center. We assume that the
black hole grew adiabatically (e.g. by accretion) to supermassive 
size early on and that a SIDM central 
density spike formed in response the hole.
We further assume that the appearance and adiabatic growth of the black hole 
took place on a timescale shorter than the SIDM relaxation
timescale, Eq.~(\ref{eq:tr_sidm}), so that the density profile in the spike 
assumed a (power-law) form,  
appropriate for a collisionless spike responding to an adiabatically 
growing black hole in a power-law halo distribution ~\cite{GonS99}. 
We then simulate below how SIDM collisions drive the
density spike to a weakly-collisional cusp around the hole 
and how heating from the cusp drives the subsequent expansion
of the halo.

\paragraph{Initial Data}. Here we adopt a simplified halo profile that
highlights the interior (cuspy) regions of an NFW halo containing a 
density spike around a central supermassive black hole.
The density profile is given by
\begin{eqnarray}
\label{eq:nfw}
\rho(r) &=& 0, \ \ \ r \leq 4M_h \ \ ({\rm capture \ region}), \\
&=& \rho_h(r_h/r)^{\gamma_{sp}}, \ \ \ \ \ \  4M_h < r \leq r_h  \ \ \
({\rm spike}), \nonumber \\
&=& \rho_h (r_h/r)^{\gamma_c}, \ \ \ \ \ r_h < r \leq R_{H}\ \ \ 
({\rm \ halo}). \nonumber
\end{eqnarray}
Defining $M_{H}$ to be the total mass of the SIDM halo, 
$R_{H}$ the halo radius and $M_h$ the mass of the black hole, 
we set the scaling parameters $M_0=M_{H}$, $R_0=R_H/25$ and
and $r_h = M_h/v_0^2$. We take $M_h/M_H = 10^{-2}$, which gives
$r_h/R_H = 4 \times 10^{-4}$. The density parameter $\rho_h$ is 
determined by substituting Eq.~(\ref{eq:nfw}) into Eq.~(\ref{eq:mass}), 
integrating over the entire SIDM halo and setting the resulting mass 
equal to $M_H$. The velocity profile is determined by substituting
Eq.~(\ref{eq:nfw}) into Eq.~(\ref{eq:hydroeq}) and integrating inward 
from the surface to find $v(r)$. 

We choose $\gamma_c = 1$, consistent with the standard NFW inner region
profile. For a spike that forms about an adiabatically growing 
supermassive black hole we then require 
$\gamma_{sp}=(9-2 \gamma_c)/(4 - \gamma_c)$~\cite{GonS99}, which
yields $\gamma_{sp}=2.33$. We set $a=4$ in the velocity-dependent 
SIDM interaction cross section as we did in Ref~\cite{ShaP14}.

We note that with the adopted initial data, $M(r_h)/M_h = 4.8 \times 10^{-5}$.
Hence the black hole greatly dominates the potential well inside 
the inner spike. In fact, given the adopted density profile, the black hole 
plays a dominant role out to $r/R_H \sim 0.1$, at which radius $M(r)=M_h$.

\paragraph{Boundary Conditions}

As we did in Section~\ref{subsub:static}, Eq.~(\ref{eq:bcinfixed}), 
we mimic the capture of
matter by the black hole by fixing a small inner radius $r_{in}$ within which
the interior SIDM mass is a vanishingly small value.
We set $r_{in}/R_H = 3.80 \times 10^{-5}$. 
At the surface we again employ exterior vacuum boundary conditions, 
Eqs.~(\ref{eq:bcout}).

\paragraph{Numerical Results}

\begin{figure}
\includegraphics[width=10cm]{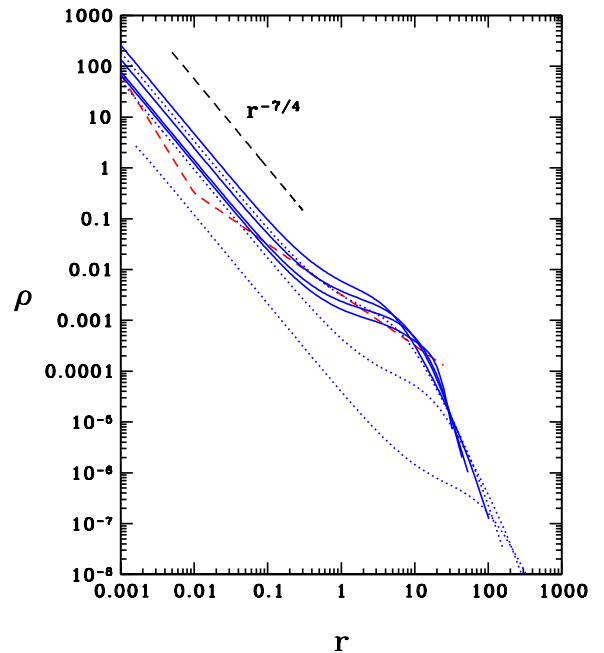}
\caption{Black hole influence on SIDM halo evolution:
snapshots of the density profile at selected times.
The lower {\it dashed} red line shows the density at time $t=0$.
Successively higher {\it solid} blue curves show the density at
$t = 18.09, 61.45, 97.95$ and $294.8$ (gravothermal collapse). 
Successively lower {\it dotted} blue curves show 
the density at $t = 539.5, 2.527 \times 10^3$
and $2.938 \times 10^4$ (re-expansion).  The upper {\it dashed} black
line shows the slope for the steady-state BW cusp solution.
All quantities are in nondimensional units (see Eq.~\ref{eq:nondim3}).
}
\label{fig:dens_BHnfw}
\end{figure}

\begin{figure}
\includegraphics[width=10cm]{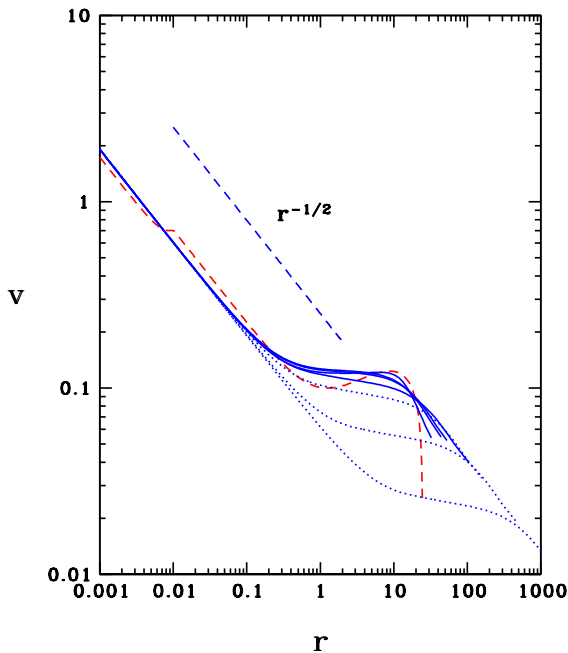}
\caption{Black hole influence on SIDM halo evolution:
snapshots of the velocity dispersion profile at the same selected times
depicted in Fig.~\ref{fig:dens_BHnfw}.
The straight {\it dashed} black
curve shows the slope for the steady-state BW cusp solution.
All quantities are in nondimensional units.
}
\label{fig:vel_BHnfw}
\end{figure}

We use 281 logarithmically spaced grid points spanning 
seven decades in M to solve the system of equations. 
The evolution equation for $s$ was integrated
implicitly using Eq.~(\ref{eq:firstlaw6}). Results are
summarized in Figs.~\ref{fig:dens_BHnfw} and ~\ref{fig:vel_BHnfw}.

Fig.~\ref{fig:dens_BHnfw} shows that early on the initial central
spike evolves to a standard weakly-collisional cusp around the
black hole. For $a=4$ the cusp exhibits the usual BW profile.
This happens early because the relaxation time is shortest 
in the cusp: $t_r(r_h)/t_{r0}=0.11$, 
where $t_r(r_h)$ is the initial relaxation
time at $r=r_h$. Shortly afterwards the cuspy NFW density 
profile tends to smooth out and develop a flatter core outside the cusp.
For the period $0 \lesssim t/t_0  \lesssim 295$ the cluster undergoes
gravothermal core collapse. For $t/t_0 \gtrsim 295$ the density in
the black hole cusp generates enough heat to eventually
reverse core collapse and drive re-expansion of the halo, as 
predicted.

The velocity dispersion shown in Fig.~\ref{fig:vel_BHnfw} quickly relaxes
to the anticipated BW solution $v^2 \approx (4/11)M_h/r$ in the BH cusp. 
The dispersion flattens out outside the black hole cusp as a
flatter, nearly isothermal density core grows around the cusp. 
As the the halo expands the velocity dispersion in the core steadily decreases in magnitude, as required by the virial 
theorem in an expanding, self-gravitating system, 
and the black hole cusp region grows in time.

\section{General Relativistic Treatment}
\label{sec:GR}

The above applications demonstrate the utility of the hydrodynamic 
conduction approximation for tracking the secular evolution of 
weakly-collisional, self-gravitating, $N$-body systems in Newtonian gravitation.
This motivates us to develop a similar approach in 
general relativity for virialized systems with  
strong gravitational fields and constituents moving 
at velocities approaching the speed of light. We previously provided such an approach to study the special case of
steady-state SIDM cusps around massive black holes in  
halo centers~\cite{ShaP14}. Here we develop the formalism 
to track the time-dependent evolution of more general, 
weakly-collisional, spherical systems. Our treatment, albeit 
approximate, is designed to fill a gap, as
as we are not aware of any other approach 
that has been employed to treat relativistic systems 
in this physical regime.

The starting point of our analysis is the metric of a 
quasistatic, spherical spacetime, which may be written as
\begin{equation}
\label{eq:metric}
ds^2 = -e^{2 \Phi} dt^2 + e^{2 \Lambda} dr^2 + r^2 d\Omega^2,
\end{equation} 
where $e^{2 \Lambda}\equiv 1/(1-2M(r)/r)$ and $M(r)$ is the 
total mass-energy of the configuration inside radius $r$.
The relativistic versions of the Newtonian 
hydrostatic equilibrium 
Eqs.~(\ref{eq:mass}) and (\ref{eq:hydroeq}) are the TOV 
equations,
\begin{equation}
\label{eq:massGR}
\frac{\partial M(r)}{\partial r} = 4 \pi r^2 \rho,
\end{equation}
\begin{equation}
\label{eq:hydroeqGR}
\frac{\partial P}{\partial r} = -(\rho+P)
  \frac{M(r) + 4 \pi r^3 P}{r(r-2M(r))}
\end{equation}
and
\begin{equation}
\label{eq:Phi}
\frac{\partial \Phi}{\partial r} = \frac{M(r) + 4 \pi r^3 P}
    {r(r-2M(r))},
\end{equation}
where $\rho$ is the total mass-energy density.
The evolution of the system is again governed by
the entropy equation, whereby Eq.~(\ref{eq:firstlaw}) 
now becomes
\begin{equation}
\label{eq:entropyGR}
\frac{d \rho}{d \tau} - \frac{\rho + P}{n} \frac{d n}{d \tau} = 
 nT\frac{ds}{d \tau} =
 -\nabla_a q^a - a_a q^a = 0,
\end{equation}
where $\tau$ is proper time,
$n$ is the proper particle number density, $T$ is the kinetic
temperature, $a^a$ is the particle four-acceleration, and $q^a$ is the heat 
flux four-vector. Here we adopt the classical Eckart formulation
of relativistic conduction~\cite{Eck40} (see also \cite{MisTW73}) 
which is adequate for illustrative purposes, 
leaving for future implementation more refined formulations that 
address the issue of noncausality and other subtleties. 
We follow our analysis in Ref~\cite{ShaP14} and
model the particles (stars or SIDM) as a perfect,
nearly collisionless, relativistic gas where at each radius 
all the particles have 
the same local speed but move isotropically. 
We may then set at each
radius $P \equiv nk_BT = \rho v^2$, where
$k_B$ is Boltzmann's constant and $v$ is the one-dimensional velocity 
dispersion measured by an observer in a static, orthonormal frame. 
We also have $\rho = \gamma \rho_0$, where $\rho_0 = mn$ is the 
rest-mass density, $m$ is the particle mass 
and $\gamma = 1/(1-3 v^2)^{1/2}$. These relations 
give $k_B T = \gamma m v^2$. Eq.~(\ref{eq:entropyGR})
may then be written as
\begin{equation}
\label{eq:entropyGR2}
\rho v^2 \frac{d }{d \tau}\ln \left[ \frac{(\gamma^2-1)^{3/2}}{\rho_0} \right]  
  = -\nabla_a q^a - a_a q^a.
\end{equation}

For a virialized system in a (quasi-)static, spherical gravitational
field the only nonzero component of $q^a$ is $q^r$, where
\begin{equation}
\label{eq:fluxGR}
q_r = -\frac{\kappa}{|g_{00}|^{1/2}} 
\frac{\partial \left(T|g_{00}|^{1/2} \right)}{\partial r},    
\end{equation}
and where $\kappa$ is the effective thermal conductivity 
and $g_{00}=-e^{2\Phi}$. We determine $\kappa$ for our
weakly-collisional (LMFP) systems by first considering the conductivity
of a relativistic, strongly-collisional (SMFP) 
gas of hard-spheres~\cite{CerK02}: 
\begin{equation}
\label{eq:condGR}
\kappa = \frac{3}{64 \pi} \frac{k_B}{\sigma_h} 
\frac{(\zeta + 5G -G^2\zeta)^2 \zeta^4 K_2(\zeta)^2}
     {(\zeta^2 + 2)K_2(2\zeta) + 5 \zeta K_3(2\zeta)}.
\end{equation}
In the above equation $\zeta = m/k_B T$, $K_n$ is a modified Bessel function
of the second kind, $G=K_3(\zeta)/K_2(\zeta)$, and $\sigma_h = d^4/4$, where
$d$ is the sphere diameter.
Next we write $\sigma_h$ in terms of the mean-free path $\lambda$,
for which 
\begin{equation}
\label{eq:lambda}
\lambda = \tau_c v_m = \frac{1}{4 \pi \sigma_h n} 
\left[ \frac{\gamma^2}{1+\gamma^2} \right]^{1/2}, 
\end{equation}
where $v_m = \sqrt{3}v$ is the mean three-dimensional speed  
and $\tau_c$ is the collision time~\cite{AndW74}. We then
substitute $\lambda$ for $\sigma_h$ 
in Eq.~(\ref{eq:condGR}), using Eq.~(\ref{eq:lambda}),  and,
following the prescription in Refs~\cite{LynE80} and ~\cite{Spi87} 
for modifying the SMFP result
to estimate the conductivity in a weakly-interacting (LMFP) gas, we 
multiply $\lambda$ by $(H/\lambda)(H/v_m t_r)$. 

In nonrelativistic (NR)
regions where $\zeta \gg 1$, $\gamma \approx 1$ and $\rho \approx \rho_0$, 
this prescription yields
\begin{equation}
\label{eq: kappaNR}
\kappa \approx \frac{75}{64}(2 \pi)^{1/2} \rho_0 v \lambda \frac{k_B}{m}
       \rightarrow \frac{75}{64}\left(\frac{2 \pi}{3}\right)^{1/2}\rho_0 
         \frac{H^2}{t_r} \frac{k_B}{m} \ \ \ \ ({\rm NR})
\end{equation}
which, together with Eq.~(\ref{eq:fluxGR}) and the Newtonian 
relations $q_r \approx L/4 \pi r^2$ and $g_{00} \approx -1$ 
leads to Eq.~(\ref{eq:flux}) for
the hard-sphere value of $b=(25/64)\sqrt{2 \pi/3}=0.565$ quoted previously.
The appropriate value of $t_r$ is given by Eq.~(\ref{eq:tr_stars}) for stars and
by Eq.~(\ref{eq:tr_sidm}) for SIDM particles. 
We again note that Ref~\cite{Spi87} adopts
$b=0.45$ as a better fit to more detailed models of Newtonian, isotropic star 
cluster evolution. We also note that we should set 
$\eta = \sqrt{6} = 2.44$ in Eq.~(\ref{eq:tr_sidm}) for 
SIDM particles moving isotropically 
at a locally constant speed $v_m$. The 
value of the scale height $H$ to assign already 
has been discussed in Section~\ref{sec:Newt}, below Eq.~(\ref{eq:flux})

In extreme relativistic (ER) regions where $\zeta \ll 1$, $v_m \rightarrow 1$ 
and $\gamma \gg 1$ we have
\begin{equation}
\label{eq:kappaER}
\kappa \approx 2 \lambda \rho_0 \frac{k_B}{m}
     \rightarrow 2 \rho_0 \frac{H^2}{t_r} \frac{k_B}{m} 
\ \ \ \ ({\rm ER})
\end{equation}
Here $t_r$ for a relativistic SIDM gas 
may be approximated by the 
collision time $\tau_c$:
\begin{equation}
\label{eq:tr_sidmER}
t_r \approx \frac{1}{\sigma \rho_0 v} 
       \left[ \frac{\gamma^2}{1+\gamma^2} \right]^{1/2} 
   \rightarrow \frac{1}{\sigma \rho_0} \ \ \ \ ({\rm ER \ SIDM}),
\end{equation}
where $\sigma$ (cross section per unit mass) 
was defined in Eq.~(\ref{eq:tr_sidm}).
The relaxation time for repeated, small-angle scattering 
for stars in a relativistic cluster is calculated
in Appendix A, and is given by
\begin{eqnarray}
\label{eq:tr_star}
t_r &\approx& \frac{3^{3/2} v^3}{8 \pi m \rho_0 \ln{(0.4N)}}
   \left( \frac{\gamma^2}{1+6 \gamma^2 v^2} \right)^2, \cr
  &\rightarrow& \frac{1}{32 \pi m \rho_0 \ln (0.4N)} 
           \ \ \ \ ({\rm ER \ stars}),
\end{eqnarray}
where $v \rightarrow 1/\sqrt{3}$ in the ER limit.

We note that the conductivity described above only takes into account 
thermal transport generated by elastic collisions between particles. 
However, there are other, 
dissipative processes that may contribute to the flux of kinetic energy.
In dense clusters of compact stars, for example, these processes 
include gravitational radiation, specifically gravitational bremsstrahlung, 
leading to the dissipative formation of binaries and their
subsequent merger~\cite{ZelP66,QuiS87,QuiS89}. Also important in 
dense stellar systems are stellar collisions and mergers, 
as well as binary heating~(see ~\cite{Spi87,QuiS90,HegA92} 
and references therein).
In SIDM halos, there also may be particle annihilation. These
dissipative processes can be especially important when the
particle velocities become relativistic, although when the cores of
virialized, large N-body
systems secularly evolve to a sufficiently
high central redshift ($\gtrsim 0.5$) they typically
become unstable to dynamical collapse, as suggested by Zel'dovich
\& Podurets~\cite{ZelP66} and demonstrated by Shapiro \& 
Teukolsky~\cite{ShaT85a,ShaT85b,ShaT86,ShaT92}(but see ~\cite{RasST89} for a  
counterexample). In any case it is possible to incorporate such  
effects by, e.g., adding appropriate heating and cooling 
terms on the right-hand side of Eq.~(\ref{eq:entropyGR}),  
but such an extension we 
shall omit in this preliminary analysis.

Evaluating Eq.~(\ref{eq:entropyGR2}) using 
$a_r = \nabla_r \ln |g_{00}|^{1/2} = \partial_r \Phi$, 
Eqs.~(\ref{eq:metric}) and (\ref{eq:fluxGR}) yield
\begin{eqnarray}
\label{eq:entropyGR3}
\rho v^2 \frac{d }{d \tau}\ln \left[\frac{(\gamma^2-1)^{3/2}}{\rho_0} \right]
&=& \frac{1}{e^{\Phi+\Lambda} r^2} 
\partial_r \left[ \kappa e^{-\Lambda}r^2 \partial_r (T e^{\Phi}) \right] \cr
&+& \frac{\kappa} {e^{\Phi+2\Lambda}} \partial_r \left( T e^\Phi \right)
\partial_r \Phi.
\end{eqnarray}
In some numerical applications it can prove helpful to employ a 
Lagrangian variable as the independent coordinate, as we did in our
Newtonian treatment. The logical choice is the 
rest-mass $M_0(r)$, where
\begin{equation}
\label{eq:restmassGR}
\frac{\partial M_0}{\partial r} = 4 \pi r^2 \rho_0 e^{\Lambda}.
\end{equation}
The resulting set of equations then becomes
\begin{equation}
\label{eq:massGR}
\frac{\partial M}{\partial M_0} = \gamma (1-2M/r)^{1/2},
\end{equation}
\begin{equation}
\label{eq:radGR}
\frac{\partial r}{\partial M_0} = \frac{\gamma (1-2M/r)^{1/2}}
       {4 \pi r^2 \rho},
\end{equation}
\begin{equation}
\label{eq:hydroeqGR}
\frac{\partial P}{\partial M_0} = -(\rho + P)
  \frac{M + 4 \pi r^3 P}{r(r-2M)}
  \frac{\partial r}{\partial M_0},
\end{equation}
\begin{equation}
\label{eq:goo}
\frac{\partial \Phi}{\partial M_0} =  
  \frac{M + 4 \pi r^3 P}{r(r-2M)}
  \frac{\partial r}{\partial M_0},
\end{equation}
\begin{eqnarray}
\rho v^2 \frac{\partial }{\partial \tau}\ln 
\left[\frac{(\gamma^2-1)^{3/2}}{\rho_0} \right] =  
\ \ \ \ \ \ \ \ \ \ \ \ \cr 
\frac{1}{e^{\Phi+\Lambda} r^2} \partial_{M_0}
\left[ \kappa e^{-\Lambda}r^2 \partial_{M_0} (T e^{\Phi})
\frac{\partial M_0}{\partial r} \right] 
\left( \frac{\partial M_0}{\partial r} \right)  \cr
+ \frac{\kappa} {e^{\Phi+2\Lambda}} 
\partial_{M_0} \left( T e^\Phi \right) \partial_{M_0} \Phi
\left(\frac{\partial M_0}{\partial r}\right)^2. \nonumber
\end{eqnarray}
The last (evolution) equation reduces to
\begin{equation}
\begin{split}
\label{eq:entropyGR4}
\rho v^2 \frac{\partial }{\partial t}\ln 
\left[\frac{(\gamma^2-1)^{3/2}}{\rho_0} \right] &=  
 \frac{4 \pi \rho}{\gamma } 
~\partial_{M_0} \left[ \frac{\kappa 4 \pi \rho r^4}{\gamma} 
\partial_{M_0}(Te^{\Phi})\right] \\
&+ \frac{\kappa(4 \pi \rho r^2)^2}{\gamma^2}
  \partial_{M_0}(T e^{\Phi}) ~\partial_{M_0} \Phi. \hspace{4cm}
\end{split}
\end{equation}
In obtaining the final form of the evolution  equation we used the relation 
$\partial_\tau \approx e^{-\Phi} \partial_t$,
which holds since the mean fluid velocity is everywhere negligible in
a virialized, spherical, quasistatic system. By implementing Eq.~(\ref{eq:entropyGR4})
the evolution advances on hypersurfaces of constant coordinate time $t$ (proper
time at infinity).

There are then seven unknowns -- $M, r, P, \Phi, \rho, \rho_0$, and $v$ 
(or $T$) --
that are determined as functions of $M_0$ by solving the five 
relations Eqs.~(\ref{eq:massGR}) -(\ref{eq:entropyGR4})
and using the two auxiliary (equation of state) relations for $P$ and $\rho_0$.
The kinetic heat flux generated by the interactions can be calculated from
\begin{equation}
\frac{L}{4 \pi r^2} = |q^a q_a|^{1/2} = |q_r| (1-2M/r)^{1/2},
\end{equation} 
using Eq.~(\ref{eq:fluxGR}).

A subset of the relativistic equations was employed in Ref~\cite{ShaP14} 
to solve for the steady-state distribution of matter in the cusp 
around a massive black
hole a the center of a weakly-collisional clusters of particles.
Included in this study were star clusters and SIDM halos. There the
central mass of the black hole
dominated the cusp and the spacetime was static Schwarzschild. 
Applications involving the full set of equations to study
clusters that secularly evolve into the relativistic regime 
are planned for the future.

\begin{acknowledgments}
It is a pleasure to thank T. Baumgarte, C. Gammie and A. Tsokaros for useful
discussions. This work has been supported in part by
NSF Grants PHY-1602536 and PHY-1662211 and NASA Grant 80NSSC17K0070 at the 
University of Illinois at Urbana-Champaign.
\end{acknowledgments}

\appendix*
\section{Relaxation Timescale for Relativistic Gravitational Encounters}

Here we provide an approximate calculation of the 
relaxation timescale due to the cumulative effect of 
multiple, small-angle, gravitational
encounters in a cluster of (point) particles moving at relativistic
speeds. We begin by treating the scattering of one test star, $m$, 
taken at rest, by another star $M$ moving at speed V relative the first. 
Since we are only interested in small-angle deflections, which are 
caused by distant encounters, 
we can take the moving star $M$ to follow a straight line trajectory
at an impact parameter $b \gg M$ from the test star. We then 
adopt the impulse approximation to determine the motion imparted
to the test star by the gravitational field of the moving star. 
We take the trajectory of the moving star to be along the $z$-axis,
$z = Vt$, and the test star to lie along the $x$-axis at
$x=b$. The impulse, imparted to the test star  
by the distant, weak field of the moving 
star, results in a velocity $\Delta v^{\perp}_m \ll 1$ 
perpendicular to the trajectory of the moving star along
the $-x$ direction. This velocity
may calculated from the Newtonian equation of motion acting on the test star:
\begin{equation}
\label{eq:eom}
\frac{d^2 x}{dt^2} = -\frac{\partial \Phi_N}{\partial x},  
\end{equation}
where $\Phi_N = - h_{00}/2$ is the Newtonian potential arising from the
moving star $M$ and $h_{00}$ is the leading order perturbation to the flat
spacetime metric, $g_{ab} = \eta_{ab} + h_{ab}$ induced by $M$. Here 
$\eta_{ab}$ is the Minkowski metric. The perturbation
$h_{a'b'}$ at the test star in a frame in which $M$ at rest is 
easily obtained from linear general relativity~(see \cite{MisTW73}, 
Exercise 18.3), 
\begin{eqnarray}
h_{0'0'} &=& h_{x'x'}=h_{y'y'}=h_{z'z'}= \frac{2M}{r'}, \cr
h_{a'b'} &=& 0, \ \ \ a' \neq b',
\end{eqnarray}
where $r'=(b^2+V^2 t'^2)^{1/2}$. The perturbation $h_{00}$ appearing
in Eq.~(\ref{eq:eom}) is then obtained from $h_{a'b'}$ above by performing a 
Lorentz boost back to the initial rest frame of the test star, 
using $t'=\gamma(t - Vz) =\gamma t$, where $\gamma = 1/(1-V^2)^{1/2}$. 
This yields
\begin{equation}
\label{eq:h00}
\Phi_N= -\frac{h_{00}}{2}=-\frac{M}{(b^2+ \gamma^2 V^2 t^2)^{1/2}}
        \left( 2 \gamma^2 V^2 +1) \right).
\end{equation}
Inserting Eq.~(\ref{eq:h00}) into Eq.~(\ref{eq:eom}) and integrating
$d^2 x/dt^2$ from $t=-\infty$ to $t=+\infty$ gives
\begin{equation}
\label{eq:delvm}
\Delta v^{\perp}_m= \frac{2M}{bV} \frac{(1+2\gamma^2 V^2)}{\gamma}.
\end{equation}

The momentum imparted to the test star along $-x$ is 
$P^{\perp}_m =\gamma_m m \Delta v^{\perp}_m \approx m \Delta v^{\perp}_m$, 
so by momentum conservation $M$ acquires a momentum
$P^{\perp}_M=\gamma M \Delta v^{\perp}_M = -P^{\perp}_m$ 
along $+x$. This gives for the velocity imparted to $M$
\begin{equation}
\label{eq:delvM}
\Delta v^{\perp}_M = \frac{2m}{bV} \frac{1 + 2 \gamma^2 V^2}{\gamma^2}.
\end{equation} 
We note that Eq.~(\ref{eq:delvM}) reduces to the correct Newtonian result
for low velocities,
\begin{equation}
\Delta v^{\perp}_M \approx \frac{2m}{bV}, \ \ \ \  V \ll 1.
\end{equation}
For high velocities Eq.~(\ref{eq:delvM}) gives
\begin{equation}
\Delta v^{\perp}_M \approx  \frac{4m}{b}, \ \ \ \ V \rightarrow 1,
\end{equation}
for which the resulting deflection angle is familiar from light bending,
\begin{equation}
\tan \alpha \approx \alpha \approx \frac{\Delta v^{\perp}_M}{V} \approx 
       \frac{4m}{b}, \ \ \ \ V \rightarrow 1.
\end{equation}

Assuming that $M$ receives repeated, randomly-oriented impulses from multiple
perturbers in time $\Delta t$, its cumulative, mean-squared perpendicular 
velocity kick becomes
\begin{eqnarray}
\langle (\Delta v^{\perp}_{M})^2 \rangle 
           &=& \sum_{i} (\Delta v^{\perp}_{M})_i^2 \cr 
           &\rightarrow& \int_{b_{min}}^{b_{max}} \left(\frac{2m}{bV}
           \frac{1+2 \gamma^2 V^2}{\gamma^2}\right)^2 dN_p  \cr
           &=& \frac{8 \pi m^2 n \Delta t}{V} 
              \ln{\left(\frac{b_{max}}{b_{min}}\right)}
              \left(\frac{1+2 \gamma^2 V^2}{\gamma^2}\right)^2,~ 
\end{eqnarray}
where $dN_p=n (V \Delta t)( 2 \pi b db)$ is the number of perturbers
and $n$ is their number density. Here $b_{max}$ is the characteristic scale of
the system, while $b_{min}$ is the impact parameter corresponding
to large-angle ($\pi/2$) scattering. The relaxation time $t_r$ can then be defined as the time 
$\Delta t$ required
for the cumulative perpendicular velocity kick to equal the initial velocity,
$\langle (\Delta v^{\perp}_M)^2 \rangle = V^2$, which gives
\begin{equation}
t_r \approx \frac{V^3}{8 \pi m \rho_0 \ln{\left(\frac{b_{max}}{b_{min}}\right)}}
   \left( \frac{\gamma^2}{1+2 \gamma^2 V^2} \right)^2.
\end{equation}
For most applications it is reasonable to approximate the logarithmic
factor as in Ref~\cite{Spi87} for Newtonian clusters: $\ln(b_{max}/b_{min}) 
\sim \ln(0.4 N)$, where N is the total number of stars. Even for
relativistic systems, we expect that $b_{min} \sim m$ and, by the virial
theorem, 
$(Nm)/b_{max} \sim V^2 \sim 1$, for which $b_{max}/b_{min} \sim N \gg 1$. 
Setting $V^2 = v_m^2 = 3v^2$ gives  
\begin{equation}
\label{eq:tr_starsGR}
t_r \approx \frac{3^{3/2} v^3}{8 \pi m \rho_0 \ln{(0.4N)}}
   \left( \frac{\gamma^2}{1+6 \gamma^2 v^2} \right)^2.
\end{equation}

We observe that in the nonrelativistic limit Eq.~(\ref{eq:tr_starsGR})
gives a relaxation time within a factor of two of 
the value quoted in Eqs.~(\ref{eq:tr_stars}) and Ref~\cite{Spi87}.
In the highly relativistic limit  Eq.~(\ref{eq:tr_starsGR}) gives a time
that scales similarly with $v$ and $\rho_0$ and is just 
a numerical factor (36) times smaller than the nonrelativistic value.

\bibliography{paper}

\end{document}